\documentstyle[prd,aps,eqsecnum,preprint,epsf,floats]{revtex}
\newif\iftightenlines\tightenlinesfalse
\tightenlines\tightenlinestrue
\def\eslt{E\llap/_T}
\def\to{\rightarrow}

\def\tb{\tilde b}

\def\tt{\tilde t}

\def\tg{\tilde g}
\def\tG{\tilde G}

\def\tw{\widetilde W}
\def\tz{\widetilde Z}

\begin{document}
\draft
\preprint{\vbox{\baselineskip=14pt%
   \rightline{FSU-HEP-081898}\break 
   \rightline{UH-511-912-98}
}}

\title{Searching for Bottom Squarks at Luminosity Upgrades of the
Fermilab Tevatron}
\author{Howard Baer$^1$, P.~G.~Mercadante$^2$, and Xerxes Tata$^2$} 
\address{
$^1$Department of Physics, 
Florida State University, 
Tallahassee, FL 32306-4350 USA
}
\address{
$^2$Department of Physics and Astronomy,
University of Hawaii,
Honolulu, HI 96822 USA
}
\maketitle
\begin{abstract}
Because of their Yukawa interactions, third generation squarks may be
substantially lighter than those of the first two generations. Assuming
that $\tb_1 \to b\tz_1$ and $\tz_1$ escapes experimental detection, we
show that experiments at the Main Injector upgrade (integrated
luminosity of 2~$fb^{-1}$) of the Tevatron should be sensitive to
$b$-squark masses up to 210 GeV for $m_{\tz_1} \leq 120$~GeV.  For
integrated luminosities of 10~$fb^{-1}$ (25~$fb^{-1}$) the sbottom mass reach
increases by 20~GeV (35~GeV). If the channel $\tb_1 \to b\tz_2$ is also
accessible, the reach becomes model-dependent and may be degraded
relative to the case where only the decay to $\tz_1$ is allowed. In
models with gaugino mass unification and $\mu$ much larger than gaugino
masses, we argue that this degradation is unlikely to be larger than
30-40~GeV.

\end{abstract}
\pacs{PACS numbers: 14.80.Cp, 14.80.Ly, 11.30.Pb}

It is well known that third generation squarks, on account of their
large Yukawa couplings, can be significantly lighter than their first
and second generation cousins, even in models where soft supersymmetry
(SUSY) breaking squark masses are universal at some scale. The large
value of the top quark Yukawa coupling has prompted several
theoretical~\cite{TOPTH} and experimental~\cite{TOPEX} investigations of
possible signals via which a light $t$-squark might be detectable at the
Tevatron, under the assumption that all other squarks as well as the
gluino are heavier. It should, however, be remembered that electroweak
gauge invariance requires that the soft SUSY breaking masses of $\tt_L$
and $\tb_L$ are identical since these are members of an $SU(2)_L$
doublet; moreover, whereas the diagonal entries in the top squark mass
squared matrix include a contribution $m_t^2$ (from electroweak symmetry
breaking), the corresponding contribution to the sbottom mass matrix is
essentially negligible. It is, therefore, not unreasonable to consider
the possibility that $\tb_1$, the lighter of the two $b$-squark
eigenstates, may be much lighter than other squarks. Moreover, in models
(such as mSUGRA) with universal squark masses at some high scale, it is
reasonable to suppose that $\tb_1 \approx \tb_L$ as long as the bottom
Yukawa coupling is small relative to its top counterpart. Experiments at
LEP have obtained lower limits of about 70~GeV on the sbottom
mass. These limits~\cite{LEP} have been obtained assuming that 
the $b$-squark mixing angle $\theta_b$ vanishes. More generally, the LEP limit
depends on $\theta_b$, and even
disappears if this angle is fine tuned so that the $Z$ exchange amplitude
exactly cancels the corresponding photon exchange. The purpose of this
paper is to examine signals via which such a light sbottom may be
searched for at future runs of the Tevatron collider, and to delineate
the reach of experiments at the Main Injector (MI) and its luminosity
upgrades that may be possible in the future.

The signals depend on how the sbottom decays. In our analysis, we assume
that the gluino is heavy so that the decay $\tb_1 \to b\tg$ is
kinematically forbidden, and further, that the lightest neutralino
$\tz_1$ is the lightest SUSY particle (LSP), so that the decay $\tb_1
\to b\tz_1$ is always allowed.
The other possible two-body decay modes are
$\tb_1 \to b\tz_i$ and $\tb_1 \to t\tw_i$. Searches at LEP 2 exclude
charginos lighter than $\sim 85$~GeV and so preclude the latter decays
unless $\tb_1$ is heavier than 260~GeV. We
will see that such heavy sbottoms are beyond the reach of the Tevatron
upgrades; it is, therefore, sufficient to focus on the case where only
the neutralino decays of $\tb_1$ are accessible.

We use ISAJET 7.37~\cite{ISAJET} for our simulation.  To model the
experimental conditions at the Tevatron, we use the toy calorimeter
simulation package ISAPLT. We simulate calorimetry covering $-4<\eta <4$
with cell size $\Delta\eta\times\Delta\phi =0.1\times 5^\circ$. We take
the hadronic (electromagnetic) energy resolution to be $50\% /\sqrt{E}$
($15\% /\sqrt{E}$).  Jets are defined as hadronic clusters with $E_T >
15$~GeV within a cone with $\Delta R=\sqrt{\Delta\eta^2 +\Delta\phi^2}
=0.7$. We require that $|\eta_j| \leq 3.5$.  Muons and electrons are
classified as isolated if they have $p_T>10$~GeV, $|\eta (\ell )|<2$,
and the visible activity within a cone of $R=0.3$ about the lepton
direction is less than 5~GeV.  For SVX tagged $b$-jets, we require a jet
(using the above jet requirement) to have in addition
$|\eta_j|<1$~\cite{FN1} and to
contain a $B$-hadron with $p_T > 15$~GeV. Then the jet is tagged as a
$b$-jet with a 50\% efficiency.

We begin by considering the simplest case, where we assume that
$m_{\tz_2} > m_{\tb_1}-m_b$ so that $\tb_1 \to b\tz_1$ with a branching
fraction of essentially 100\%. In this case, the signal naively consists
of two $b$-jets recoiling against $\eslt$ from the two neutralinos that
escape detection. The dominant Standard Model backgrounds come from $W +
j$, $Z \to \nu\nu +j$, $Z \to \tau\tau +j$ and $t\bar{t}$ production. We
consider the signal to be observable above background if, for a given
integrated luminosity, ({\it i})~the signal exceeds ``5$\sigma$''; {\it
i.e.} $S \geq 5\sqrt{B}$, where $S$ and $B$
are the expected number of signal and background events, respectively, and
({\it ii})~$S \geq 0.2B$. We also require a minimum of 5 signal events.
To enhance the signal relative to the SM background,
we impose the following requirements, hereafter
referred to as the basic cuts:

\begin{enumerate}
\item  at least two jets with $p_T(j_1)>30$~GeV, $p_T(j_2)> 20$~GeV;
\item at least one jet in $|\eta_j|<1$;
\item $\eslt > 50$~GeV;
\item $\delta\phi(\vec{\eslt},\vec{p}_{Tj})>30^\circ$;
\item for di-jet events only, $\delta\phi(\vec{p}_{Tj1},\vec{p}_{Tj2})
<150^\circ$;
\item at least one SVX tagged B;
\item no isolated leptons ($e$ or $\mu$).
\end{enumerate}

The SM backgrounds at a 2~TeV $p\bar{p}$ collider, with these basic cuts, are
shown in the first row of Table 1. We use CTEQ3L
structure functions~\cite{CTEQ} for all our calculations, and take
$m_t=175$~GeV. Except in the case
of the $t\bar{t}$ background the tagged $b$-jet comes from QCD
radiation. For an integrated luminosity of 2~$fb^{-1}$ expected to be
accumulated at the MI upgrade of the Tevatron, the $5\sigma$ level falls
just short of our $S \geq 0.2B$ requirement. To see how to further
enhance the signal,
we note that $t\bar{t}$ events form the dominant background. 
Because of the lepton veto, much of this background comes when
one of the tops decays into a tau lepton that decays
hadronically, while the other top decays completely hadronically. These
events are, therefore, likely to have large jet multiplicity, in
contrast to the signal (as well as the other backgrounds in the
Table). We are, therefore, led to
impose the additional requirement,
\begin{enumerate}
\setcounter{enumi}{7}
\item $n_j =2,3$
\end{enumerate}
designed to further reduce the top background with relatively modest
loss of signal. The corresponding background levels are shown in the
second row of Table~1. The entry +8 in the first column denotes the cuts
over and above the basic cuts 1-7. Indeed, we see that the top
background is reduced by a factor 5, and no longer dominates.

Turning to the signal, we use ISAJET to generate $\tb_1\bar{\tb}_1$ events
where $\tb_1 \to b\tz_1$, and pass these through our simulation with the
cuts discussed above. With these assumptions, the cross section is
completely determined by $m_{\tb_1}$ and $m_{\tz_1}$. 
The reach of a 2~TeV $p\bar{p}$ collider is
shown in the $m_{\tb_1}-m_{\tz_1}$ plane in Fig.~1. 
The
diagonal solid line marks the boundary of the region where $m_{\tb_1}
\geq m_{\tz_1} + m_b$. With our assumptions, we are required to be below
this line, since otherwise, $\tb_1$ would be the LSP.
The dot-dashed contour shows the reach that should be attainable
(using cuts 1-8) with
an integrated luminosity of 2~$fb^{-1}$ corresponding to
the anticipated data sample expected at the MI. The dotted lines denote
signal cross sections after these cuts. 

It is possible that larger data samples might be accumulated after
several years of MI operation, or at TeV33, a proposed~\cite{TEV33}
luminosity upgrade of the Tevatron. Here, we analyse the $b$-squark
reach for integrated luminosities of 10~$fb^{-1}$ and 25~$fb^{-1}$. The
background, which now dominantly comes from $Z \to \nu\nu + j$ events, has
to be further reduced in order to satisfy our $S/B$ requirement.
It is reasonable to suppose that both the radiated jets 
that recoil against the high in $p_T$ $Z$ boson
these events are frequently on the opposite side as the $Z$ in the
transverse plane. We are thus led to impose, in addition to cuts 1-8, a
further requirement,
\begin{enumerate}
\setcounter{enumi}{8}
\item $\Delta\phi{j_1, j_2} \geq 90^\circ$,
\end{enumerate} 
which significantly reduces the vector boson backgrounds, as can be seen
from the third row in Table~1. For an integrated luminosity of
10~$fb^{-1}$, the $5\sigma$ level is just below the 20\% requirement of
15.4~$fb$ which we take to be the observability level for this case.
This is shown as the dashed contour in Fig.~1. 

For the yet higher integrated luminosity of 25~$fb^{-1}$, the high event
rate makes it possible
to require double $b$-tagging, 
\begin{enumerate}
\setcounter{enumi}{9}
\item $n_b \geq 2$,
\end{enumerate}
to greatly reduce the vector boson background. As
before, the top
quark background is reduced by cut 8 on the jet multiplicity. The
background levels corresponding to just the $n_b \geq 2$ cut (over and
above the basic cuts), as well as
this cut in conjunction with the cut on jet multiplicity, are shown in
the next two rows of Table~1. Note that even after these stringent cuts,
the $5\sigma$ cross section which
determines the observable level implies that there should be more than 100
signal events in the data sample. The corresponding reach is shown as
the solid contour in Fig.~1.

Several comments are in order.
\begin{itemize}
\item The reach of the MI extends to $m_{\tb_1}=210$~GeV as long as the
LSP is not too close in mass to $\tb_1$; for $70$~GeV~$\leq m_{\tb_1}
\leq 160$~GeV, the signal is observable for $m_{\tz_1} \alt
0.8m_{\tb_1}$. For heavier $\tb_1$ the signal is observable up to
200~GeV as long as the LSP is lighter than 135~GeV. This result is
model-independent as long as $\tb_1 \to b\tz_1$ and $\tz_1$ escapes
detection in the experimental apparatus.

\item With the cuts suggested, typically $\geq 100$ signal events are
expected in the region where we claim an observable signal, for all
three integrated luminosity cases studied here.

\item There is a considerable increase in the region of the
$m_{\tb_1}-m_{\tz_1}$ plane that can be probed with larger integrated
luminosity. Assuming that the detectors and other systems function the
same way in a high luminosity environment, a data sample of  25~$fb^{-1}$
would enable us to detect $\tb_1$ as heavy as 240~GeV for an LSP up to
160~GeV.

\item If $m_{\tz_1}$ is close to $m_{\tb_1}$, the solid contour is
actually slightly inside the dashed one. This is because the $b$-jet then tends
to be soft so that the efficiency for double
$b$-tagging is considerably reduced. 

\item In addition to SVX tagging, the CDF collaboration had
used~\cite{TOP} soft lepton tags (SLT) to enhance the $b$ tagging
efficiency for its top quark analysis. For the $b$-squark signal
discussed in this paper, we found that attempting to include lepton tags
leads to a much bigger increase in the background relative to the signal
and {\it actually degrades the reach}. This is because $W + j$ events
where the lepton from a $W$ is accidently within the jet (which is then
tagged as $b$) are a significant source of ``fake'' background
events. This is not so for the top quark search since, in that case, in
addition to the lepton to tag the $b$, another isolated lepton was
required to be present.

\item Finally, we stress that efficient SVX tagging of $b$-jets is
crucial for this search. Our analysis may be somewhat optimistic in that
we have not included the possibility that charm or light quark or gluon
jets might fake the $b$ in our estimate of the SM backgrounds. While we
do not expect these backgrounds to be overwhelmingly large, the reader
can use the cross section contours in the figure to obtain a rough idea
of how the reach would be altered at Run~II once these backgrounds are
included, or for that matter, if the $b$ tagging efficiency turns out to
be different from the assumed 50\%.
\end{itemize}

We should note that it may be possible to modify our cuts (for $\tb_1$
search at the MI) to search for $b$-squarks in the CDF Run I data
sample. Because the SVX tagging efficiency seemed to be
evolving~\cite{TOP} as the run progressed, we have made no attempt to
project the sbottom mass range that the CDF collaboration might be able
to probe using the data that they have already collected.

Up to now we have assumed that $\tb_1 \to b\tz_1$ and further that
$\tz_1$ escapes detection. It is reasonable to ask what happens if this
is not the case: in particular, can the reach be substantially degraded
from that shown in Fig.~1? We confine ourselves to models where
$R$-parity is conserved, and where $\tz_1$ is the (stable) LSP, noting
only that the reach may be considerably degraded if $\tz_1$ decayed
hadronically via $\tz_1 \to qqq$ (and its conjugate mode) via $R$-parity
violating interactions. Since LEP constraints already preclude the decay
$\tb_1 \to t\tw_1$, we are led to consider the possibility that $\tb_1
\to b\tz_2$ is also allowed. The signal now depends on the branching
fraction for this decay, as well as the decay pattern of $\tz_2$. In
other words, the signal depends both on the $b$-squark mixing angle
as well as on the parameters of the neutralino mass matrix.

We have already argued that in many models it is reasonable to suppose
that $\tb_1 \approx \tb_L$. To make our analysis tractable, we will
assume that this is the case. We will further assume that $|\mu|$ is
much larger than electroweak gaugino masses. Assuming gaugino mass
unification, the two lighter neutralinos are approximately the
hypercharge gaugino and the $SU(2)$-gaugino, with $m_{\tz_2} \approx
m_{\tw_1} \approx 2m_{\tz_1}$. Note that these assumptions fix the
branching fraction for $\tb_1 \to b\tz_{1,2}$ decays in terms of the
sparticle masses. It is also worth pointing out that if $\tz_2 \simeq
SU(2)$-gaugino, it essentially decouples from $\tb_R$, so that the maximum
impact of the $\tb_1 \to b\tz_2$ indeed occurs when $\tb_1 =\tb_L$.

Since we are interested in seeing how much the reach of the Tevatron may
be reduced from that shown in Fig.~1, we consider extreme limits for how
$\tz_2$ might decay \cite{FN2}. If $\tz_2$ dominantly decays to leptons
via $\tz_2 \to \ell\bar{\ell}\tz_1$, sbottom pair production would
result in characteristic $bb+4\ell +\eslt$ and $bb+2\ell + jets +\eslt$
events for which the background is small, and the corresponding reach,
presumably, larger than that in Fig.~1. If $\tz_2 \to b\bar{b}\tz_1$ it
may be possible to reduce the background (and hence increase the reach)
by requiring two (or more) $b$-tags. The worst case ``realistic''
scenario is when $\tz_2$ decays into jets which are not amenable to any
tagging \cite{FN3}.  To simulate this situation, we have forced $\tz_2$
to decay via $\tz_2 \to u\bar{u}\tz_1$ and run these events through our
simulation, and once again obtained the reach for the three choices of
integrated luminosity in Fig.~1.

The results of our analysis for this case are illustrated in Fig.~2.  The
upper diagonal line is as in Fig.~1, while the lower line is where
$m_{\tb_1} =2m_{\tz_1} +m_b \simeq m_{\tz_2}+m_b$. In our analysis, we
have adjusted $A_b$ to cancel the off diagonal term in the sbottom mass
matrix in order to make $\tb_1 =\tb_L$. We have fixed $\mu = 500$~GeV
and $\tan\beta =2$; this value of $\mu$ is large enough for $\tz_1$ and
$\tz_2$ to be gauginos to a very good approximation. When the decay
$\tb_1 \to b\tz_2$ is inaccessible, the reach should be as given by our
analysis above; {\it i.e.}, the reach illustrated by the dot-dashed,
dashed and solid contours until just above this line, is identical to
that in Fig.~1. The contours below this line are obtained as described
shortly, and show the extent to which the reach might be reduced if the
$\tb_1$ can also decay to $\tz_2$.  These contours in Fig.~2 turn
inwards just slightly above this line precisely because the relation
$m_{\tz_2} = 2m_{\tz_1}$ is slightly violated by our finite choice of
$\mu$.

As before, the cuts have to be optimized for each value of integrated
luminosity. Cut 8, which was so effective in reducing the $t\bar{t}$
background in our analysis in Fig.~1 now leads to too large a signal
loss because of the additional jets from $\tz_2$ decay. For the case of
2~$fb^{-1}$ we found that the best strategy was to simply use the basic
cuts 1-7 for which the 20\% signal level is just a bit above the
$5\sigma$ level. The dot-dashed contour in the Fig.~2 shows the
corresponding reach for the MI. We see that even in this ``worst case''
scenario, the MI reach is diminished by no more than 25-30~GeV. 

For an integrated luminosity of 10~$fb^{-1}$, the analysis is more
complicated. Just below the lower diagonal line, the jet multiplicity is
still not large because the $b$-jet from $\tb_1 \to b\tz_2$ tends to be
soft and/or the branching fraction for the decay is still not large: in
this case, the additional cuts 8+9 are still effective and the signal
(presumably from events where at least one of the sbottoms decays
directly to $\tz_1$) remains observable above background. However, for
larger values of $m_{\tb_1}-m_{\tz_2}$, too much of the signal is
eliminated by the jet multiplicity cut, and the reach in $m_{\tb_1}$ is
considerably reduced. In this case, however, it is possible to obtain an
observable signal by requiring double $b$-tagging, without any restriction
on jet multiplicity. For an integrated luminosity of 10~$fb^{-1}$, we
thus consider the signal to be observable if {\it either} it is
observable using the additional cuts 8 and 9 (the signal is then still
background-limited), {\it or} it is observable using the additional cut
10. The boundary of the plane that can be probed with 10~$fb^{-1}$ is
shown by the dashed contour in Fig.~2, where the kink merely reflects
that this region is a composite of two such regions as we have just
described.

Finally, for the 25~$fb^{-1}$ case, as before we consider double-tagged
events, for which the $t\bar{t}$ background dominates. Since the top
background sample (after the basic cuts) is expected to contain a pair
of jets from the decay of a $W$ boson, we further require,
\begin{enumerate}
\setcounter{enumi}{10}
\item $m_{j_1, j_2} \leq 60$~GeV, where $j_1$ and $j_2$ are the two
highest $p_T$ untagged jets in the event. If an event has less than two
untagged jets, we retain it as part of the signal.
\end{enumerate}
The solid contour in Fig.~2, shows the reach for an integrated
luminosity of 25~$fb^{-1}$ with the additional cuts 10 and 11, for which
the background level is just 23.3~$fb$. We see that the reach may be
diminished by about 40~GeV from that shown in Fig.~1.  As with our
earlier analysis, with our new cuts designed to pull out the signal when
the $\tb_1 \to b\tz_2$ might be accessible, we expect $\geq 100$ signal
events (for all three cases of integrated luminosity) over the entire
region where the signal should be observable. We also stress that the
degradation of the reach by 30-40 GeV that we have found is the largest
it could be in a wide class of models. Most models will lead to a
smaller reduction in the reach.  In fact, in some scenarios, {\it e.g.}
gauge-mediated models~\cite{GMM} where $\tz_1 \to \gamma \tG$, or
$R$-parity violating models~\cite{RVIOL} where $\tz_1 \to
\ell\bar{\ell'}\nu$, the reach may even be larger than that shown in
Fig.~1, because the presence of hard photons or leptons serves to
greatly reduce the SM background.

While this may be obvious, it may be worth emphasizing that since we do
not {\it a priori} know whether $\tb_1 \to b\tz_2$ is accessible, two
separate analyses might be needed. If no signal is found with the
analysis where it was assumed $\tb_1 \to b\tz_1$, this means that either
($m_{\tb_1}, m_{\tz_1}$) is outside the observable region in Fig.~1,
{\it or} there is a sbottom signal with $\tb_1 \to b\tz_2$ which is
being effectively cut out, for instance, by the $n_j=2,3$ cut (cut 8) used in
this analysis. To ensure that one is not throwing the baby out with the
bathwater, it is thus important to reanalyse the data using the
alternative set of cuts, specifically designed to retain the signal in
this case.

We have also examined the case where $\tz_2$ produced with exactly the
same branching fraction as in Fig.~2 decays invisibly. In this case,
there is a sharp reduction in the reach just below the kinematic
boundary for the $\tb_1 \to b\tz_2$ decay (since the daughter $b$
frequently fails to pass the cut on the second jet) but for lowest
values of $m_{\tz_1}$ in the figures, the reach contours lie within
10~GeV of the corresponding $m_{\tb_1}$ values in Fig.~1. Except for a
tiny region very close to the $\tb_1 \to b\tz_2$ decay boundary, the
reach is more degraded in the case where $\tz_2 \to u\bar{u}\tz_1$ than
in the case where $\tz_2$ decays invisibly.

In summary, motivated by the fact that the $b$-squark may be lighter
than other squarks we have attempted to assess the capability of the MI
and possible future luminosity upgrades of the Tevatron
to identify a SUSY signal from $\tb_1$
pair production. Assuming that $\tb_1 \to b\tz_1$ and that $\tz_1$
escapes detection, we have shown that
it should be possible for experiments at the MI to detect $b$-squark
signals over SM backgrounds for $m_{\tb_1} \leq 210$~GeV, even if the
LSP is quite heavy. The capability of tagging $b$-jets in the central
region with high efficiency and purity is crucial for this
detection. Luminosity upgrades to 10~$fb^{-1}$ (25~$fb^{-1}$) should
increase the reach to $\sim 230$~GeV ($\sim 245$~GeV). The reach may be
somewhat degraded if sbottom can also decay into $\tz_2$. 
We have argued
that in many models (with $\tz_1$ as a stable LSP) this degradation is
typically smaller than 30-40~GeV, but could be larger if $\tz_1$ is
unstable and decays only to hadrons.

\begin{center}
\begin{table}[ ]
\caption[]{Standard Model background cross sections in $fb$ to the
$b$-squark signal after the basic cuts 1-7 described in the text, as
well as after additional cuts designed to further reduce
backgrounds. The ``plus entries'' in the first column refer to the cuts
in addition to the basic cuts; for instance, the last row has cuts 1-8
together with cut 11. We take $m_t=175$ GeV.}

\bigskip
\begin{tabular}{|l|c|c|c|c|c|}
CUT &$W+j$ & $Z \to \nu\nu +j$ & $Z \to \tau\tau +j$ & $t\bar{t}$ & $Total$ \\ 
\tableline
Basic & 65.5 & 92.4 &2.6  & 195 & 356 \\
$+8$ & 51.6 & 80.6 & 2.1 & 36.7 & 172 \\
$+8+9$ &21.9 & 26.2 & 0.9  & 28.0 & 77 \\
$+10$ & 5.6 & 7.2 & 0.1 & 37.5 & 50.4 \\
$+8+10$ & 3.8 & 6.6 & 0.1 & 6.7 & 17.2 \\
$+8+11$ &5.0 & 6.6 & 0 & 11.7 & 23.3

\end{tabular}
\end{table}
\end{center}

%
\acknowledgments 
P.M. was partially supported by Funda\c{c}\~ao de
Amparo \`a Pesquisa do Estado de S\~ao Paulo (FAPESP).  This work was
supported in part by the U.~S. Department of Energy under contract
numbers DE-FG-FG03-94ER40833 and DE-FG02-97ER41022.

%
%

\newpage
%

\centerline{\bf \large{Figure Captions}}
\begin{description}

\item[Fig.\ 1]
The region of the $m_{\tb_1}-m_{\tz_1}$ plane that can be probed at a
2~TeV $p\bar{p}$ collider, assuming that $\tb_1 \to b\tz_1$ and that
$\tz_1$ escapes detection. The sbottom signal should be detectable 
with the observability criteria defined in the text in the region below
the dot-dashed, dashed and solid contours for an integrated luminosity of
2~$fb^{-1}$, 10~$fb^{-1}$ and 25~$fb^{-1}$, respectively, with the set
of cuts given in the text. These cuts have been optimized for each value
of integrated luminosity.
Also shown in the figure are contours of constant
signal cross section after cuts 1-8 for the 2~$fb^{-1}$ case. The 34~$fb$
contour marks the $0.2B$ level that we require as a minimum for the
signal. The diagonal line
marks the boundary of the region beyond which $m_{\tb_1} > m_b+m_{\tz_1}$.
\item[Fig.\ 2] The same as Fig.~1 except that the decay $\tb_1 \to
b\tz_2$ is also allowed when kinematically accessible below the lower
diagonal line.  To compute the branching fraction for this decay, we
assume that $\tb_1 =\tb_L$, and that $\tz_1$ and $\tz_2$ are essentially
hypercharge and $SU(2)$ gauginos, respectively, as discussed in the
text, and that $m_{\tz_2} \approx 2m_{\tz_1}$. To illustrate the largest
degradation of the reach in this scenario, we assume that $\tz_2$ always
decays via $\tz_2 \to u\bar{u} \tz_1$. The dotted lines are again
contours of fixed cross section after the basic cuts 1-7 we use for the
MI analysis. The dot-dashed contour that denotes our projection of the
MI reach also corresponds to $S=0.2B$.

\end{description}



\begin{references}
\bibitem{FN1} In Run II of the Tevatron, we understand that $b$-tagging
may be possible~\cite{ENO} out to $|\eta_b|=1.6-2$. We have checked
that this does not help to extend the reach that we have obtained. We
have, therefore, used the more conservative range for
$b$-tagging.
\bibitem{FN2} In
the extreme case where the branching fraction for $\tb_1 \to b\tz_2$ is
100\% when this decay is allowed, and if $\tz_2$ only decays invisibly,
the reach can once again be read off from Fig.~1.
\bibitem{FN3} We expect that this leads to a lower $\eslt$ than in
the case when it decays invisbly. We will return to this issue shortly.
\end{references}
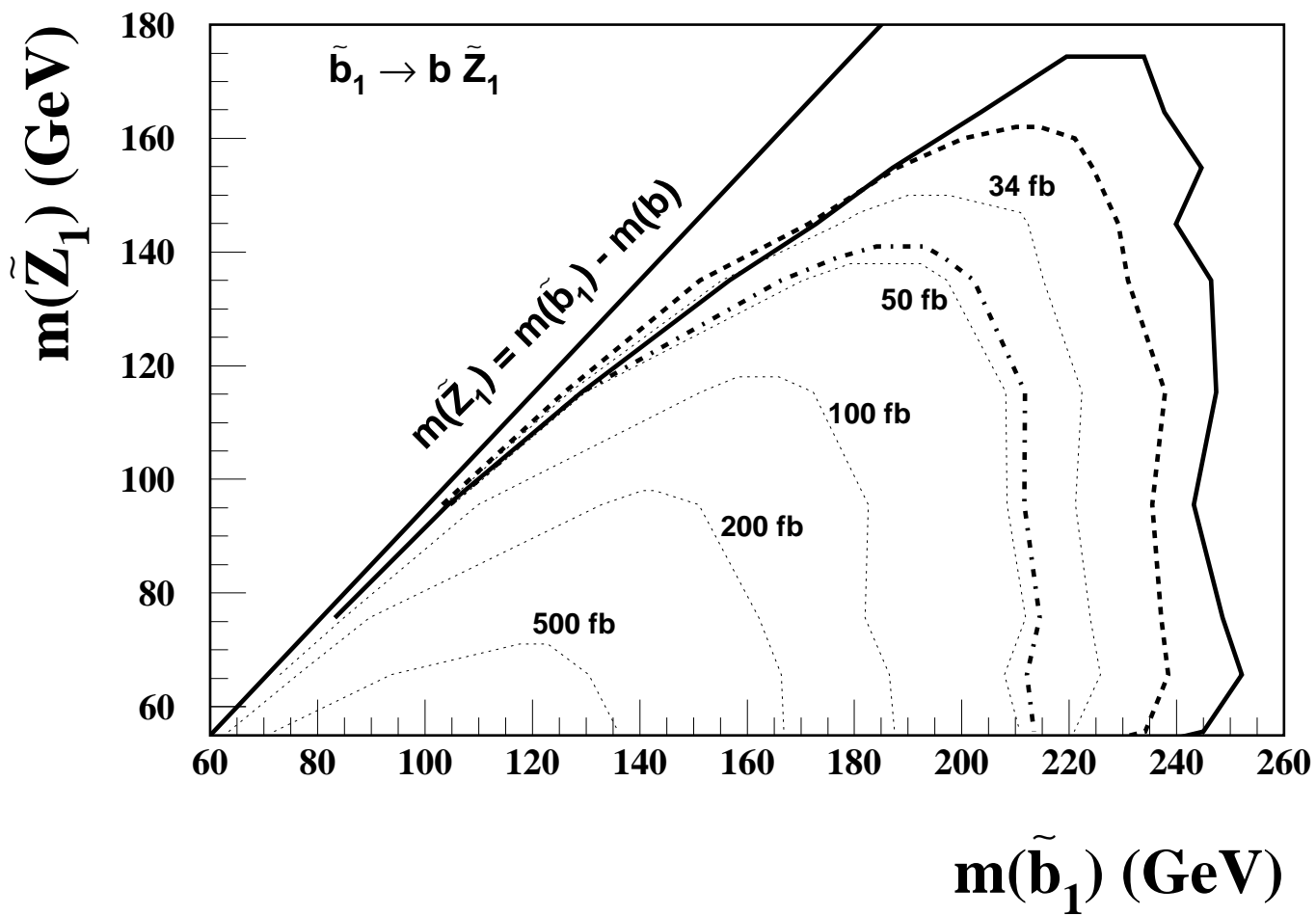
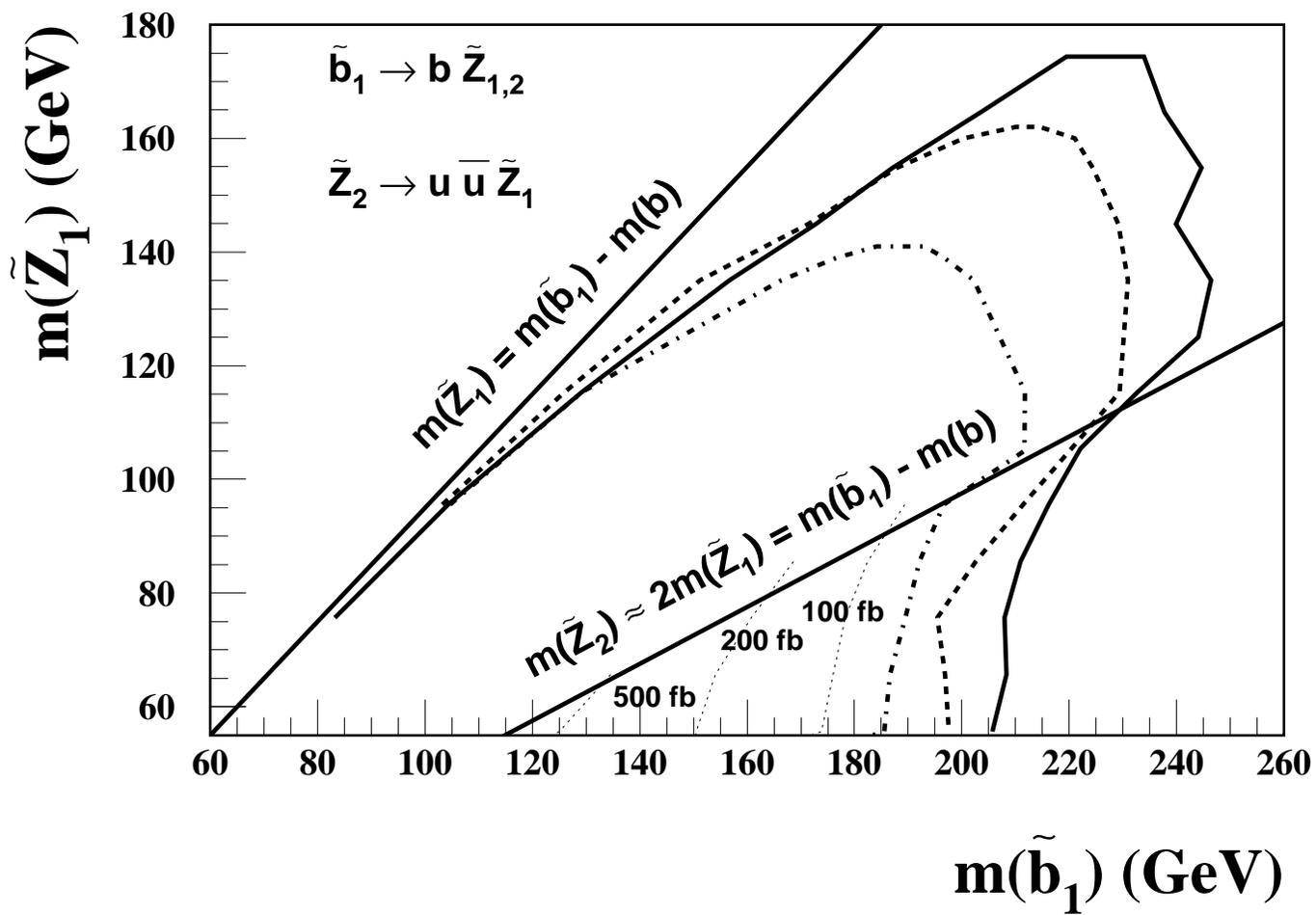
\end{document}